\documentclass[12pt,a4paper]{article}
\usepackage[margin=1in]{geometry}
\usepackage{amssymb,amsmath,amsthm,url}
\usepackage[noblocks]{authblk}
\usepackage{setspace}
\usepackage{epigraph}
\usepackage[square]{natbib}
\newtheorem{thm}{Theorem}

\newtheorem{defn}{Definition}

\makeatletter
\newcommand{\specificthanks}[1]{\@fnsymbol{#1}}
\makeatother
\date{}
\title{A measure of authorship by publications}
\author[1]{Conan Mukherjee\thanks{{\tt Email:conan.mukherjee@gmail.com}}}
\author[2]{Ranojoy Basu\thanks{{\tt  Email:ranojoy.basu@iimu.ac.in}}}
\author[3]{Aftab Alam\thanks{{\tt Email:aftab@phy.iitb.ac.in}}}
\affil[1]{Economics group, Indian Institute of Management Calcutta, India}
\affil[2]{Department of Economics and Finance, Indian Institute of Management Udaipur, India}
\affil[3]{Department of Physics, Indian Institute of Technology Bombay, India}

\begin{document}
\maketitle
\begin{abstract}
Measuring publication success of a researcher is a complicated task as publications are often co-authored by multiple authors, and so, require comparison of solo publications with joint publications. In this paper, like \cite{price1981multiple}, we argue for an egalitarian perspective in accomplishing this task.

More specifically, we justify the need for an ethical perspective in quantifying academic author by identifying certain ethical difficulties of some popular contemporary indices used for this purpose. And then we show that for any given dataset of research papers, the unique method satisfying the ethical notions of {\it identity independence} and {\it performance invariance} must be the egaliatarian $E$-index proposed by \cite{bps} and \cite{price1981multiple}. In our setting, this egalitarian method divides authorship of joint projects equally among authors and sums across all publications of each author.

\vspace{3mm}
\end{abstract}
\pagebreak
\section{Introduction}
\epigraph{
``You Cannot Manage What You Cannot Measure"
}{\textit{W. Edwards Deming}}
In academic organizations and research laboratories, it is often required to measure and compare the cumulative impact and research performance of individuals. Such comparisons are mostly used to {compare authors and undertake important professional decisions} like hiring, promotion, granting tenure, awarding grants etc.  However, {there is a lack of consensus on how to quantify the success of an academic on the basis of accomplishments in teaching, grant writing and academic publication.}

{In this paper, we focus on the particular issue of academic publication, and provide an intuitive method with strong justifications to {\it measure} the publishing success of an author from a given dataset of papers. In particular, we explicitly recognize academic publishing as the cooperative exercise that it generally is, and use a cooperative game theoretic framework to arrive at a measure of the success that can be attributed to an individual given any list of her joint or solo author projects. Interestingly, unlike most of the related literature, our proposed method allows us to trade off the academic worth of a single author publication vis-a-vis a joint author publication.

We call our method $E${\it -index} as it measures the academic success of an author in an {\it egalitarian} manner, by equally dividing credit for all her joint projects and then summing across all her publications in the dataset. More importantly, we show that this index is a strong embodiment of certain desirable properties or {\it axioms} that, in our opinion, any index evaluating authorship should satisfy. In particular, it is the {\it only} index that satisfies the axioms of {\it identity independence}, and {\it performance invariance}.}

In our model, the notion of: (i) identity independence requires that {the authorship ascribed to an author} should not depend on her own identity;  and (ii) performance invariance requires that increase in publication quality or quantity, should enhance the academic credit assigned to a researcher. We feel that these properties must necessarily be satisfied by any reasonable method measuring authorship. Any identity dependent method would lead authors getting compared on individual characteristics other than their academic contributions, which would lead to formation of prejudiced author comparisons. A non-monotonic method would allow loss of academic credit upon increase in academic contribution, and thus, create perverse incentives for academic research. {Finally, a wasteful index that does not exhaust sum of all citations in a dataset, would fail to utilise all the relevant  information embedded in the data, and hence, imply an incorrect evaluation of an author's academic performance.}

In section \ref{hindex}, we start by (i) noting {the limitations} inherent in popular measurement notions of $g$-index and $h$-index that are widely used in several academic disciplines, and (ii) describing how our $E$-index addresses these concerns. Then, in section \ref{litreview}, we present the comparison to $E$-index with some notable recent results. Section \ref{model} presents the notations and a detailed discussion of the axioms used. Section \ref{res} presents the main result and a limited empirical example contrasting the performance of $E$-index with $h$-index in the field of Economics. Section \ref{disc} discusses a few philosophical issues pertaining to our approach of quantifying academic authorship. Finally, section \ref{conc} concludes our paper.

\subsection{Relation to $h$-index and $g$-index.}\label{hindex}
Over the last decade, $h$-index proposed by \cite{hirsch2005index}, has emerged as the most popular and accepted measure of academic authorship. It is defined as the largest number $h$ of an author's publications that have at least $h$ citations. In fact, as reported in \cite{egghe2010hirsch}, within two years of its introduction, $h$-index was incorporated by Scopus and Web of Science as a  publication success indicator. Further, \cite{ball2007achievement} reports (on page 737) that
\begin{quote}\it
``{\rm [$h$-index]}is becoming widely used informally, for example to rank applicants for research posts''.
\end{quote}
{ This popularity of $h$-index is not surprising given the elegant and simple manner in which it measures publication success of an author.}

{Yet, for all its wide acceptance, we feel that $h$-index might not be able to accommodate all the available information, in particular, the extent of citations generated by each publication.} Indeed, \cite{egghe2006theory} and \cite{egghe2010hirsch} mention that $h$-index combines number of papers (an indicator of quantity) and corresponding citations (an indicator of quality or impact) in a manner that is {\it insensitive} to one or several outstanding highly cited papers. To note the complications that may arise due to this property of $h$-index: consider a scenario where author A has $3$ {\it solo} publications of $100$ citations each, and author B has $4$ {\it joint} author publications of $4$ citations each. Then A has an $h$-index of $3$ which is {\it less} than that of B (who has $h$-index of $4$). Thus, while any serious academic will recognize A as a far more successful researcher than B, $h$-index would state otherwise.

To circumvent this shortcoming with $h$-index, \cite{egghe2006theory} came up with a simple variant called $g$-index. Given a set of articles ranked in decreasing order of the number of citations that they received, the $g$-index is the largest number such that the top $g$-articles received together at least $g^{2}$ citations. For example, a $g$-index of $10$ for an academic means that the {\it group} of top $10$ cited papers that she has written, has been cited at least $10^{2}=100$ times. It is easy to see that $g$-index allows highly cited papers to bolster low-cited papers, for any list of publications by a researcher, $g\geq h$.

{However, both these measures limit an academic's publication success to be no more than the number of papers written by her}. To see this consider the scenario where author A has $4$ {\it solo} publications of $100$ citations each, and author B has $6$ {\it joint} author publications of $6$ citations each. Then A has a identical $h$ and $g$-index value of $4$ while author B has an identical $h$ and $g$ index value of $6$. Thus, $B$ gets identified as a better researcher than A by both indices, in spite of having produced (in joint efforts with others) far less popular work than A (who has written only solo papers) - {\it merely} because she has written a {\it larger} number of papers. This tendency to reward quantity, in our opinion, an issue of concern, as in general, the more profound results require a longer time to accomplish, and so, pursuing such  path-breaking results may mean forgoing several less ambitious papers that are relatively lower hanging fruits. This, in turn, would stall development and expansion of a knowledge generation within a discipline. As we argue later, our $E$-index does not bound measure of academic authorship by number of papers written.

Another limitation with both these indices is that they don't take into account the difference between solo author and co-author publications. This feature entirely overlooks any need for proration of authorship in joint publications. However, this lack of proration can lead to rampant unjustified co-authorships, and sometimes, false authorship.\footnote{\cite{stephan1996economics} presents a concept of {\it false} authorship that results out of an agreement between a group of authors to share credits of different papers with the understanding that the heavy lifting for different papers will be done, not by all, but a few members of this group. Given the unbridled increase in co-authorship over years, it is difficult to establish absence of such practices.} In fact, proliferation of such undesirable authorship has now forced several fields to explicitly formalize the definition of an author.\footnote{For example, the {American Physical Society} and the {International Council of Biomedical Journal Editors} have established specific rules for authorship (see Section VII in ~\cite{liebowitz2014willful}). These definitions are available at the webpages, \url{https://www.aps.org/policy/statements/02_2.cfm} and \url{https://www.councilscienceeditors.org/resource-library/editorial-policies/white-paper-on-publication-ethics/2-2-authorship-and-authorship-responsibilities}, respectively.} Unfortunately, despite such drastic measures, \cite{tarnow2002coauthorship} argues that there is widespread suspicion of undesirable authorships contravening these definitions. In an attempt to discourage such tendencies, \cite{berk1989irresponsible}, then an editor of a leading radiology journal, called for devaluing the impact of co-author papers on academic authorship. Our $E$-index follows this advice by suitably devaluing authorship generated out of joint papers in an ethical manner.

{Finally, both these indices return whole numbers as values. This creates hindrance in implementation when value of a publication is measured by the impact factor of the journal where it appeared.} This is specially true during tenure decisions of young faculties who have not had enough time for their citations to reasonably accumulate. Note that such impact factors, even for the best journal, widely vary across disciplines. In fact, they may very well take some positive fraction value. Therefore, in the extreme, if all papers of an author are published in such journals with fractional impact factors, she will get assigned the $h$-index (or $g$-index) equal to $0$, {\it irrespective} of the number of papers she may have published. Undoubtedly, this would be very unfair, and also, cause a loss in discriminatory power among applicants. Our $E$-index can easily be adapted to different measures of academic authorship without entailing these difficulties.

\subsection{Relation to other literature}
\subsubsection*{Theoretical literature}\label{litreview}
As argued in \cite{perry2016count}, we believe that any index must have axiomatic foundations. And so, unlike \cite{hirsch2005index}, we begin with certain desirable axioms and then look for the index that is characterized by them. {Hence, this paper is distinct from \cite{marchant2009axiomatic}, \cite{bouyssou2014axiomatic}, \cite{quesada2011axiomatics}, \cite{woeginger2008axiomatic2}, \cite{woeginger2008axiomatic1}, which start with a pre-existing index function and then, identify properties that completely characterize this function.

The papers most relevant to the present study are \cite{chambers2014scholarly}, \cite{perry2016count}, \cite{marchant2009score}, \cite{liebowitz2014willful}, \cite{bruno2014economics}, and \cite{szwagrzak2015co}. The first two papers adopt an individual author perspective, and so, do not deal with the issue of splitting successes of joint publications.\footnote{ In an earlier working paper \cite{perry2014count}, they propose a method of capturing in their index, the information on number of co-authors in a paper, and provide a suitable adjustment to their axiomatized ordinal index. Their adjusted index for an author's {\it list} of $n$ publications is $\left[ \sum_{l=1}^n \frac{x_l^\sigma}{k_l} \right]^{\frac{1}{\sigma}}$ where $x_l$ denotes the citation earned, $k_{l}$ the number of authors respectively of the $l^{th}$ paper and $\sigma>1$ is a parameter. Note that like our paper, they deflate the citations of a paper by a function of the number of its authors and when $\sigma=2$ the index can be shown to be identical to ours by suitable monotonic transformation (provided worth of each paper is measured by square of it's citations). However, unlike their work, our proposed method of deflation is obtained as an implication of our axioms.} \cite{marchant2009score} allows for division of author credit in joint papers and provides a characterization (that is independent to ours) that calls for equal division of authorship in such cases.\footnote{This result appears as Theorem 5 in \cite{marchant2009score}.} \cite{liebowitz2014willful}, too, discusses proration of authorship and shows that if author credits are not properly prorated, there can be excessive co-authorship. However, he assumes that coauthors are equally productive and does not present any axiomatic characterization. \cite{bruno2014economics} shows in a strategic model that the lack of attention paid to number of co-authors while measuring research productivity is an implication of the prevailing market pay structure. In contrast to these papers, we address the issue of proration of author credit explicitly, and use standard axioms of resource allocation to provide complete characterizations.

\cite{szwagrzak2015co} study a similar problem in the following manner. They characterize a method of splitting authorship of joint publications, and use a weighted sum of these individual contributions to construct an index which they call {\it Co-score}. Unfortunately, this interesting theoretical approach poses two crucial difficulties in practical usage: (i) their Co-score is difficult to compute because it has no explicit formula in general cases, and (ii) the axiomatic foundation underlying Co-score assumes that all authors in the dataset have {\it some} positive citation for any {\it one} of their solo works.\footnote{This makes it problematic to apply their score on datasets where there is an author who has no solo publication, but many highly cited joint author publications.} In contrast, our proposed index is obtained as implication of reasonable axioms that are universally applicable, and has explicit formula that can be easily computed.

Finally, we must mention that the idea of dividing the worth of a paper by the number of authors, to obtain the extent of authorship can attributed to her, was first proposed by \cite{price1981multiple}. This idea was further investigated and justified in a strategic team performance setting by \cite{bps}.}

\subsubsection*{Empirical literature}

As shown by \cite{kuld2018rise}, there has been a dramatic increase in co-authorship in Economics as discipline over the period 1996 - 2014.\footnote{\cite{kuld2018rise} conclude that there is a correlation between co-authorship and career stage, that is, young authors tend to publish more single authored papers.} Accordingly, there has been a wide array of empirical research on exploration of the relationship between co-authorship and extent of academic success (as observed by citations, publications, conference presentations, and so on).\footnote{\cite{hamermesh2018citations}, \cite{sauer1988estimates}, \cite{hilmer2015fame}, \cite{berger2016does}; \cite{garfield1999journal}; \cite{hamermesh1982scholarship}; \cite{smart1996citation} etc.} An interesting strand of this literature focuses on whether citations of an article affect salaries regardless of the number of co-authors or not. The underlying premise of this research question is that higher professional success in terms of papers cited should lead to higher salaries given that every additional citation to a co-authored paper leads to an equal increase in aggregate citation count of all authors. According to \cite{hamermesh2018citations}, the first article to examine this issue was \cite{sauer1988estimates}, who found that the returns to citations of a two-authored article were almost exactly half of those to citations of a solo-authored paper, after controlling for covariates like the number, length and journal level etc. This result provides support to our proposed idea of equal proration of author credit from co-authored papers.\footnote{ Recently however \cite{hilmer2015fame} examines a similar issue and finds that salary implication of citation counts does not appear to be discounted for co-authorship at all.}

Of course, using salaries as a proxy for individual productivity may be criticized on the grounds of the unrealistic assumption of a perfectly functioning competitive market of academicians (where both administrator and professors are price takers). However, in absence of reliable data on individual productivity, one has to resort to some sort of proxy to capture variation on this dimension.\footnote{While there has been a robust debate on what should be a proper metric to measure contribution of a paper, in recent times, a substantial amount of academic decision making procedures relating to hiring, promotion, grant etc. prefer to use citations as an appropriate measure of intellectual content of a paper. In fact, as reported by \cite{ellison2013does}, relevant authorities in United Kingdom have recently announced their intention to transform the Research Assessment Exercise (RAE) from a process that relies on peer reviews - to one that is based on bibliometric data.} This is where our result assumes significance for empirical research. We show axiomatically that, {\it no matter what metric} an empirical researcher chooses to measure contribution of a paper -  individual productivity of an academician A for any given dataset of papers (which can be as large as possible), is simply sum of the {\it average} values of papers authored by A in the data set. \footnote{\cite{hollis2001co}, \cite{ductor2015does} uses similar averaging of value of papers to measure individual productivity.}

Note that theoretically, our measure of individual productivity is ambivalent about desirability of the ever-increasing propensity to co-author papers. That is, our idea of averaging contribution of a paper allows for incentives to reap honest intellectual gains from academic collaborations. Nobel laureate George J. Sigler describes these gains in \cite{stigler2003memoirs} as:\footnote{See page 36.}
\begin{quote}
\it Association with a group of able colleagues is a strong advantage that a professor usually has over a nonacademic economist. Frequent exchanges with strong minds and powerful scientific imaginations that have
a deep understanding of the problems one is struggling with are invaluable in discovering errors and eliminating strange perspectives that creep into one’s work''.
\end{quote}

Interestingly, the general empirical findings on relationship between co-authorship and individual productivity reflect similar ambivalence on desirability of co-authorship. For example, \cite{ductor2015does} shows that co-authorship leads to a positive effect on individual productivity after taking into consideration the endogeniety inherent in the co-authorship; \cite{laband2000intellectual} documents a higher probability of acceptance for co-authored papers compared to solo authored papers; \cite{chung2009relation} reports that the average quality measured by citation frequency is higher for co-authored papers.

On the other hand, \cite{medoff2003collaboration} shows that academic productivity is not significantly affected by collaboration after controlling for article length, journal and author quality, and subject area; \cite{mcdowell1992effect} do not find any significant relationship between co-authorship and academic productivity using cross-sectional data on academics\footnote{\cite{mcdowell1992effect} regressed  the number of articles produced by an individual (with co-authored articles discounted by the number of authors) on the percentage co-authored.}; \cite{hollis2001co} finds that co-authorship leads to lower academic productivity\footnote{\cite{hollis2001co} measure individual productivity as the sum of published pages weighted by quality indices and discounted by the number of authors.}, and as mentioned earlier \cite{liebowitz2014willful} argue that current levels of co-authorship are in excess of what is desirable.

 Given a lack of consensus on the actual direction of relationship between individual productivity and co-authorship coupled with the continuous increase in co-authorship, our paper may be used to augment the empirical research by providing axiomatic foundations to the data measuring of individual productivity, and hence, provide a clearer idea of social desirability of academic co-authorship.

\section{Model}\label{model}
 Fix a set of authors $N=\{1,2,\ldots,n\}$ where $n\geq 2$ and define $\rho(N)$ to be the set of non-empty subsets of $N$. Let $P=\{p_1, p_2, \ldots, p_k\}$ an arbitrary set of papers whose authors are members of $N$. For each paper $p_t \in P$, define $v(p_t) \geq 0$ to be the value of the paper measured by some pre-defined metric.\footnote{The most basic of such metrics could the simple count of citations of a paper.} Define a dataset to be the pair $(P, v)$ where $v = (v(p_t))_{t=1}^k $. For any $t=1,\ldots,k$, let $A(p_{t})$ be the set of authors of paper $p_{t}$. Note that, by construction, for all $t$, $A(p_t) \in \rho(N)$, and $\cup_{t=1}^{K} A(p_t)= N$.

{As mentioned earlier, we take explicit cognizance of the fact that academic research publications are mostly cooperative exercises, and hence, use cooperative game theoretic techniques to analyze a dataset of publications. Hence, we use the information presented in a dataset to arrive at an estimate of {\it aggregate} academic success that can be attributed to different groups of authors. Note that in a given dataset $(P,v)$, some subgroups of $N$ may have written no papers, or some subgroups may have written more than one papers. The most obvious way of accomplishing this accounting exercise would be to take simple sum of citations earned by all papers written by any subgroup $S \subseteq N$, use it to generate a cooperative game $(N, c)$ where $c:\rho(N) \mapsto \mathbb{R}_+$ such that for all $S \in \rho(N)$, $c(S):= \sum\limits_{p\in P: A(p)=S} v(p)$. In the present context of measuring authorship, we call this characteristic function $c(.)$ of the generated cooperative game, the {credit function} {\it induced by dataset}.\footnote{Note that it would not be impossible to provide an axiomatic justification of using a simple sum of citations to generate such a credit function. However, we feel this would be an unnecessary exercise given the wide acceptability, in the field of Economics, of using sums to capture aggregate quantities like joint profits, total resource endowment etc.}

Note that our objective is to provide an index measuring the publishing success of authors involved in {\it any possible} dataset $(P,v)$. Since, {\it a priori}, the dataset, which we eventually get to work with, may be any one of {\it infinitely} many possible datasets: the credit function $c(.)$ can be visualized as a point in $\mathbb{R}_+^{\rho(N)}$. And so, the necessity of a measurement index being effective in comparing authors, no matter what dataset is chosen, implies that such index be a mapping $\phi : \mathbb{R}_+^{\rho(N)} \mapsto \mathbb{R}^{N}$. Further, for any citation $c \in  \mathbb{R}_+^{\rho(N)}$ that may be induced by the chosen dataset, the sum of index values ascribed to authors in $N$, $\sum_{i\in N} \phi_i(c)$, must exhaust the total publishing success embodied in the dataset, $\sum_{p \in P} v(p)$ (or else we would be ignoring information in dataset that is relevant to author comparisons). For simplicity of notation, henceforth, we denote  $\mathbb{R}_+^{\rho(N)}$ by $\mathcal{C}$. Thus, any author comparison index must belong to the following functional space:
\[
\left\{f:\mathcal{C} \mapsto \mathbb{R}^N_+ :  \sum_{i\in N} f_i(c) = \sum_{S \in \rho(N)} c(S) \right\}
\]
}

\vspace{3mm}
Note that any such index measuring academic success of authors must not discriminate on lines of identities markers such as gender, race etc. That is, an index must be {identity invariant} in the following manner:
\begin{defn}
$\phi$ satisfies identity independence (II) iff for all bijections $\pi :N \mapsto N$, all $c \in \mathcal{C}$, and all $i \in N$,
$$\phi_{\pi(i)}(\pi c)=\phi_i(c)$$
where $\pi c(\pi (S))= c(S)$ for all $S \in \rho(N)$.
\end{defn}

\noindent
{Note that this definition of identity independence is identical to the definition of symmetry in the seminal paper by \cite{shapley1953value}.

\vspace{3mm}
 Further, any index measuring academic success of an author must depend only on the papers that she has authored in solo or joint capacity. This property reflects the philosophical idea presented in \cite{bl}, where acquiring of knowledge is conceptualized as a community enterprise. A direct implication of this conceptualization is that an increase in success of an academic A should not detract from the credits ascribed to some other academic B. More precisely, any index must be { performance invariant} so that measure of academic success ascribed to an author should only be a function of the papers that she has written.
\begin{defn}
$\phi$ satisfies {\it performance invariance} (PI) iff for all $c,d \in \mathcal{C}$, and all $i\in N$,
\[
[\forall\: S\subseteq N\setminus\{i\}, c(S\cup  \{i\}) =   d(S\cup \{i\})] \Longrightarrow \phi_i(c) = \phi_i(d)
\]
\end{defn}
}

\vspace{3mm}
{Finally, we formally define the index proposed by \cite{bps} to measure authorship: $E$(galitarian)-{\it index}. It shares authorship of every group of  authors equally and then sums across the publications of each author in the dataset, to obtain her index value.}
\begin{defn}
For any dataset $(P,v)$, the authorship ascribed by $E$-index to any $i\in N$ is,
\[
\sum\limits_{p\in P: i\in A(p)} \frac{v(p)}{|A(p)|}
\]

where $v(p)$ is the value of paper $p \in P$ and $|A(p)|$ is the set of authors of paper $p$.\footnote{The index $\phi(\cdot)$ is continuous in $c$ which we think is a desirable property.}
\end{defn}

\section{Result}\label{res}
The following theorem states our main result. It shows that if one accepts identity independence, performance invariance and as the necessary properties that every index claiming to measure an author's publication success must satisfy, then the only option available is $E$-index.

\begin{thm}\label{TH:egalitarian}
An index satisfies II and PI if and only if, it is $E$-index.
\end{thm}

\noindent{\bf Proof:}

\noindent
{\it Proof of Sufficiency.} We first note that for any dataset $(P,v)$, {\bf (a)} $\sum\limits_{p\in P: i\in A(p)} \frac{v(p)}{|A(p)|} = \sum\limits_{S \subseteq N: i\in S} \frac{c(S)}{|S|}$, where $c(.)$ is the induced credit function. Define for any $c \in \mathcal{C}$, $\phi^e_i(c):= \sum\limits_{S \subseteq N: i\in S} \frac{c(S)}{|S|}$, and observe that for any credit function $c \in \mathcal{C}$ that may get induced by $(P,v)$, it can easily be seen that {\bf (a)} implies that $\phi^e_i(c)$ satisfies II and PI. This establishes the sufficiency of the result.

\vspace{4mm}
\noindent
{\it Proof of Necessity.} We need to show that any index $\phi(.)$ satisfying II and PI, must ascribe to an author $i\in N$ the authorship value $\sum\limits_{S \subseteq N: i\in S} \frac{c(S)}{|S|}$. That is, we must establish that for all possible characteristic functions $c \in \mathcal{C}$ that may potentially get generated from dataset $(P,v)$, $\phi_i(c) = \phi^e_i(c), \forall \; i\in N$. This is accomplished below using the technique of induction.

\vspace{2mm}
Consider any credit function $c \in \mathcal{C}$ such that $c(S)=0$ for all $S\in \rho(N)$. By II, for all $i\neq j$, $\phi_i(c)=\phi_j(c)$, and so, implies that $\phi_i(c)=0$ for all $i\in N$. Define a function $\kappa :\mathcal{C} \mapsto \mathbb{N}$ such that $\kappa(c):= |\rho(N)| - |\{S \in \rho(N): c(S)=0\}| + 1$. Fix a $k \in \{1, \ldots, |\rho(N)|\}$ and suppose that for all $c \in \mathcal{C}$ such that $\kappa(c) \leq k$, $\phi_i(c)= \sum\limits_{S \subseteq N: i\in S} \frac{c(S)}{|S|}$ for all $i\in N$. In the following paragraphs we show how our supposition (or, the  induction hypothesis); implies that for all $c \in \mathcal{C}$ such that $\kappa(c) = k+1$, $\phi_i(c)= \sum\limits_{S \subseteq N: i\in S} \frac{c(S)}{|S|}$ for all $i$.

Fix a $c\in \mathcal{C}$ such that $\kappa({c})=k+1$ and define the set $\hat{N}:=\{i\in N: \exists \; S\subseteq N\setminus \{i\} \mbox{ such that } S \neq \emptyset \mbox{ and } c(S) \neq 0\}$. Therefore, $N \setminus \hat{N}$ is the set of agents $i$, for the given credit function $c$, such that the academic contribution of any group of authors not containing $i$ is $0$. We call these agents in $N\setminus \hat{N}$ as the {\it star} authors in $c$. If all authors are star authors, that is $\hat{N} =\emptyset$, then $c(S)=0$ whenever $S\neq N$; and so, no group of authors other than the grand coalition can produce a positive academic contribution.  Therefore, it easily follows that, by II, $\phi_i(c)=\phi_j(c)$ for all $i\neq j \in N$. And so, by construction for all $i\in N$, $\phi_i(c)= \frac{c(N)}{n} = \sum\limits_{S \subseteq N: i\in S} \frac{c(S)}{|S|}$.

Consider the other possibility where all authors are not stars, that is, $\hat{N} \neq \emptyset$. For any $i\in \hat{N}$, construct a credit function $c_i \in \mathcal{C}$ such that; $c_i(S^i) =0 \neq c(S^i)$ for some $S^i \in \rho(N)$ with $i\notin S^i$, and $c_i(S)=c(S)$ for all $S \neq S^i$. By construction of $\hat{N}$, credit function $c_i$ is well defined. Further, $\kappa(c_i) = k$ and so, by induction hypothesis, $\phi_i(c_i) = \sum\limits_{S\subseteq N: i\in S} \frac{c_i(S)}{|S|} = \sum\limits_{S\subseteq N: i\in S} \frac{c(S)}{|S|}$ for all $i\in \hat{N}$. {Therefore, by PI, $\phi_i(c) = \phi_i(c_i) = \phi_i^e(c)$.}

Now, if $\hat{N}=N$ implying that there are no star authors, then the result follows trivially. If $\hat{N} \subset N$, that is, the set of star authors $N\setminus \hat{N}$ is non-empty; then by construction, for any $S \in \rho (N)$, $c(S)>0$ only if $[N \setminus \hat{N}] \subseteq S$. By applying II for all bijections $\pi: N \mapsto N$ such that $\pi (i) =i$ for all $i \in \hat{N}$, and $\pi (N\setminus \hat{N}) =N\setminus \hat{N}$, we get that $\phi_i(c)=\phi_j(c)$ for all $i\neq j \in N \setminus\hat{N}$. Therefore, by construction, we get that; for any $i\in N\setminus \hat{N}$, we get that
\[
\begin{array}{lcl}
\phi_i(c)&=&\frac{1}{n - |\hat{N}|}\left\{ \sum\limits_{S \in \rho(N)} c(S) -  \sum\limits_{i\in \hat{N}} \phi_i^e(c)\right\} \\
&&\\
&=& \frac{\sum\limits_{i\in N\setminus \hat{N}} \; \sum\limits_{S\subseteq N: i\in S}\frac{c(S)}{|S|}}{n - |\hat{N}|}
\end{array}
\]
By construction, $N\setminus \hat{N} \neq \emptyset$ implies that for all $i\neq j \in N\setminus \hat{N}$, $\{S \subseteq N: i\in S, c(S)>0\} = \{S \subseteq N: j\in S, c(S)>0\}$ and so, the right hand side of the equation above collapses to $\sum\limits_{S\subseteq N: i\in S}\frac{c(S)}{|S|}$. Thus, the result follows. \qed

\vspace{5mm}
{Now, as mentioned earlier, one of our objectives in this paper is to provide an alternative to the $h$-index. And so, it would be interesting to see, in some measure, how differently does the $E$-index perform with actual data vis-a-vis the $h$-index. To get a sense of this comparison, we present the following (very simple) example that reflects the contrast in these two indices.}
\subsection{Example}
\label{ex2}
\nonumber
We present an example which shows that rankings generated by $E$-index and $h$-index are likely to vary widely, and hence, suggests that the information ignored while computation of $h$-index may lead to inappropriate rankings. To make our point, we focus on six senior superstar academics in the age group 75 - 81 whose emphatic careers have shaped the very discipline of Economics (and Finance): Robert J. Barro, Robert F. Engle, Eugene Fama, James Heckman, Michael C. Jensen, Robert Lucas Jr., Robert C. Merton, Joseph E. Stiglitz. The advantage of considering such senior researchers is that their research finding have had substantial time to be disseminated widely leading to a comprehensive maturing of respective citations. Of course, the downside of this choice is that we would have to have a large dataset of papers encompassing their lives' works, and compute the $E$-index values for a large number of authors that have co-authorship connections to them.

Fortunately, a great amount of ranking data is freely available on Economic academicians on the {\it Ideas} bibliographic online database.\footnote{\url{https://ideas.repec.org/}} On this website two specific rankings are of great relevance with respect to our paper. The first is a ranking of top 5\% authors according to their $h$-index values ``{\it as of July 2019}''.\footnote{\url{https://ideas.repec.org/top/top.person.hindex.html}. Accessed on 14th August, 2019.}, while the second is a ranking of the ranking of top 5\% authors according to the sum of ``their papers' citations divided by corresponding authors - ``{\it as of July 2019}''.\footnote{\url{https://ideas.repec.org/top/top.person.anbcites.html} and \url{https://ideas.repec.org/top/top.person.anbcites.html#explain}. Accessed on 14th August, 2019.} The latter uses the same formula to quantify authorship as our $E$-index.

Thus, if we consider our initial dataset to be the set of all published (English) papers in Economics, the second of the aforementioned rankings gives us the rank of these authors among all other Economists in terms of $E$-index. We compare these ranks with the ranks accorded by $h$-index to these acclaimed researchers.

\[
\begin{array}{|ccccc|}
\hline
\mbox{\it Author }& &\mbox{ $E$ } && \mbox{ $h$ }\\ \hline
	\mbox{\small Robert J. Barro }&& 1 && 7\\
	\mbox{\small James Heckman}&& 2&&2\\
	\mbox{\small Eugene Fama} && 3 && 174\\
\mbox{\small Robert Lucas Jr.}&& 5&&89\\
\mbox{\small Robert F. Engle } && 6 && 17\\
\mbox{\small Joseph E. Stiglitz}&& 7&&3\\
\mbox{\small Michael C. Jensen}&& 14&&336\\
\mbox{\small Robert C. Merton}&& 15&&336\\
\hline
\end{array}
\]

There are two points we would like to make with respect to the table above. First, a simple glance across the rankings suggests wide variation in ranking numbers.\footnote{Unfortunately, computation of rank correlation coefficients would be very difficult given the large number of authors who have published in Economics. However, a cursory glance at the variation in ranks of a Nobel laureate like Eugene Fama suggests that there is likely be very little correlation between the two methods of measuring authorship.} Second, beyond a threshold rank value, $h$-index loses significant discriminatory power. For example, the ranking on $h$-index on aforementioned Ideas website the rank 336 is assigned to more than $50$ authors. This appears to be another implication of ignoring relevant information of citation data of papers, and may lead to difficulties in applying $h$-index in practical policy issues.

\section{Discussion}\label{disc}
\subsection{Axioms}
The technical contribution of this paper is to establish that E-index is the only possible evaluation procedure that respects the axioms of identity independence (II) and performance invariance (PI). Hence, to any administrator, our E-index is only as acceptable as our II and PI axioms. In this subsection of our paper, we present some additional discussion on value of our axioms from normative as well as positive perspectives.

We begin with the II axiom. It appeals to the primordial idea of symmetry as the basis for beauty/desirability in every sphere of human existence since time immemorial.\footnote{Indeed, as Aristotle mentions in {\it Metaphysics}:
\begin{quote}
\it “The chief forms of beauty are order and symmetry and definiteness, which the mathematical sciences demonstrate in a special degree.”(See \cite{sart})
\end{quote}} In our setting, this axiom implies that any evaluation procedure must be symmetric with respect of academic authorship: that is, if any two agents switch identities to exchange their academic performances, their evaluations should also get interchanged. From a positive perspective, any evaluation procedure that fails to satisfy this property is likely to lead to policy decisions susceptible to litigation on grounds of discrimination. The PI axiom, on the other hand, presents the strategic advantage of providing appropriate incentives to career progression through high quality publication. On a normative level, as mentioned earlier, this property reinforces the collaborative and harmonious spirit of academics as a profession.

Of course, along with the aforementioned axioms of II and PI, an administrator may want to impose several other kinds of behaviour restriction or axioms on the chosen evaluation method. Our characterization result, however, presents an unfortunate reality that the only possible method that satisfies these two axioms is our $E$-index. So if there is an extra property P that an administrator may want to impose (along with II and PI) such that $E$-index fails to satisfy P: then our result implies that there cannot be {\it any} evaluation procedure that satisfies all three properties. In other words, if the $E$-index is not agreeable to an administrator who happens to believe in the II and PI axioms, there will be {\it no other} evaluation procedure that would meet the needs of the administrator.

Note that we do not claim $E$-index will have the requisite strategic impact to eliminate free riding or false authorship. However, $E$-index does limit the benefits of such free riding by devaluing success of a paper by number of authors. Moreover, our method of devaluation is unlikely to deter honest research collaborations aiming to solve problems of great social or disciplinary importance, simply because, these research outputs are likely to be receive large enough citations to compensate for number of authors.

On a general strategic note, the exact pattern and contribution of authorship is a complicated private information on which all authors may not agree. Even if this lack of consensus is ignored, a strategic incomplete information model set in the Harsanyi paradigm is likely to have multiple equilibria. More importantly, such models presume complicated higher order belief structures that is often difficult to motivate when there is great amount of heterogeneity among players (in this, administrators and authors). On the contrary, an axiomatic approach to the problem such as ours, provides sharper policy prescription that rely on clear normative foundations.

\section{Other Issues}
In this paper we treat authorship as a cardinal quantity that is `{\it interpersonally comparable}' both in levels and gains. That is, we believe that differences in index numbers across authors have intuitive meaning. This allows us to conceptualize proration of academic credit for joint papers. In contrast, one may argue that academic authorship is too abstract a notion to allow for any sort of proration. However, as noted in \cite{liebowitz2014willful}, without sufficient proration to identify individual contribution in joint papers, there could be excessive co-authorship.\footnote{\cite{liebowitz2014willful} demonstrates by a statistical exercise; how increase in co-authorship over the years can be better explained by incomplete proration than by increased specialization.} Therefore, {in the interest of preserving the sanctity of knowledge generation,} we believe that it is better for any discipline to treat authorship as a cardinal variable that can be prorated. {As mentioned earlier, this sentiment is reflected in \cite{berk1989irresponsible} and \cite{price1981multiple} too.}

Further, our analysis compares authors only on basis of the {metric of success (citation number or} some other relevant measure of contribution) of their publications, accommodatingly suitably {for the cases where there are multiple author projects}. However, for many such joint author papers, the order in which names of coauthors appear in print often is {\it publicly} believed to be the agreed order of contribution to the paper. It may be argued that this additional information should affect the method of proration of authorship in joint author papers, and thus, affect any index that ranks authors.

We do not disagree with the merit of this argument. Ideally, a method of ranking authors should depend on {metrics of publication success} as well as the order in which names of authors are displayed. However, this is a complicated problem as one would have to combine a set of cardinal numbers with a set of ordinal ranks to obtain an index or ranking method. Further, there are many disciplines like social sciences, where journals are publicly known to report the names of authors in an alphabetical manner, thereby leading to a situation where the aforementioned information simply does not exist.  More importantly, \cite{hirsch2005index} himself ignores this issue of accounting for the displayed order of author contributions altogether. Therefore, considering the popularity that $h$-index has enjoyed among administrators over years, we feel that this issue is not a major hindrance to practical use of our index, at least as long as a better index accomplishing the aforementioned exercise is not made available to the academic community.

Finally, there lies a great potential for practical use in our approach of measuring academic performance on the basis of datasets of research articles. This approach allows us to answer a variety of interesting questions like: who is the best performing researcher in a department, or who is the best young researcher with at least one paper in XYZ journal, or who is the best female (or male) researcher over last five years etc. Arriving at satisfactory answers to such questions would be necessary to formulate ethical and effective administrative policies in academia.

\section{Conclusion}\label{conc}
In this paper, we note that the popular measures of academic authorship like $h$-index and $g$-index may yield counter-intuitive and unfair comparisons among researchers. We {propose an alternative method of measuring authorship from a given dataset of publications by investigating all possible ways of accomplishing this task that} satisfy two basic properties of identity independence and performance invariance. We find that the {\it unique} method that satisfies these properties is the intuitive $E$-index.
\section{Acknowledgement}
We are grateful to Professors Philip Reny, Tommy Andersson, Debashish Bhattacherjee, Satya R. Chakravarty, Sugata Marjit and Manipushpak Mitra for their valuable advice. Discussions with Jens Gudmundsson and Professor D. J. Saikia are also gratefully acknowledged. A substantial part of this paper was developed while the first author was visiting the Department of Economics, Lund University. Their kind hospitality is greatly acknowledged. Errors are ours. Aftab Alam acknowledges the financial support from IIT Bombay via the SEED grant project code $13IRCCSG020$. 



\bibliographystyle{apalike}

\bibliography{citation}

\begin{thebibliography}{}

\bibitem[Ball, 2007]{ball2007achievement}
Ball, P. (2007).
\newblock Achievement index climbs the ranks.
\newblock {\em Nature Publishing Group}.

\bibitem[Berger, 2016]{berger2016does}
Berger, J. (2016).
\newblock Does presentation order impact choice after delay?
\newblock {\em Topics in cognitive science}, 8(3):670--684.

\bibitem[Berk, 1989]{berk1989irresponsible}
Berk, R.~N. (1989).
\newblock Irresponsible coauthorship.
\newblock {\em American Journal of Roentgenology}, 152(4):719--720.

\bibitem[Blais, 1987]{bl}
Blais, M. (1987).
\newblock Epistemic tit for tat.
\newblock {\em The Journal of Philosophy}, 84:363--375.

\bibitem[Bose et~al., 2010]{bps}
Bose, A., Pal, D., and Sappington, D. (2010).
\newblock Equal pay for unequal work: Limiting sabotage in teams.
\newblock {\em Journal of Economics \& Management Strategy}, 19:25--53.

\bibitem[Bouyssou and Marchant, 2014]{bouyssou2014axiomatic}
Bouyssou, D. and Marchant, T. (2014).
\newblock An axiomatic approach to bibliometric rankings and indices.
\newblock {\em Journal of Informetrics}, 8(3):449--477.

\bibitem[Bruno, 2014]{bruno2014economics}
Bruno, B. (2014).
\newblock Economics of co-authorship.
\newblock {\em Economic Analysis and Policy}, 44(2):212--220.

\bibitem[Chambers and Miller, 2014]{chambers2014scholarly}
Chambers, C.~P. and Miller, A.~D. (2014).
\newblock Scholarly influence.
\newblock {\em Journal of Economic Theory}, 151:571--583.

\bibitem[Chung et~al., 2009]{chung2009relation}
Chung, K.~H., Cox, R.~A., and Kim, K.~A. (2009).
\newblock On the relation between intellectual collaboration and intellectual
  output: Evidence from the finance academe.
\newblock {\em The Quarterly Review of Economics and Finance}, 49(3):893--916.

\bibitem[Ductor, 2015]{ductor2015does}
Ductor, L. (2015).
\newblock Does co-authorship lead to higher academic productivity?
\newblock {\em Oxford Bulletin of Economics and Statistics}, 77(3):385--407.

\bibitem[Egghe, 2006]{egghe2006theory}
Egghe, L. (2006).
\newblock Theory and practise of the g-index.
\newblock {\em Scientometrics}, 69(1):131--152.

\bibitem[Egghe, 2010]{egghe2010hirsch}
Egghe, L. (2010).
\newblock The hirsch index and related impact measures.
\newblock {\em Annual review of information science and technology},
  44(1):65--114.

\bibitem[Ellison, 2013]{ellison2013does}
Ellison, G. (2013).
\newblock How does the market use citation data? the hirsch index in economics.
\newblock {\em American Economic Journal: Applied Economics}, 5(3):63--90.

\bibitem[Garfield, 1999]{garfield1999journal}
Garfield, E. (1999).
\newblock Journal impact factor: a brief review.

\bibitem[Hamermesh, 2018]{hamermesh2018citations}
Hamermesh, D.~S. (2018).
\newblock Citations in economics: Measurement, uses, and impacts.
\newblock {\em Journal of Economic Literature}, 56(1):115--56.

\bibitem[Hamermesh et~al., 1982]{hamermesh1982scholarship}
Hamermesh, D.~S., Johnson, G.~E., and Weisbrod, B.~A. (1982).
\newblock Scholarship, citations and salaries: Economic rewards in economics.
\newblock {\em Southern Economic Journal}, 49(2).

\bibitem[Hilmer et~al., 2015]{hilmer2015fame}
Hilmer, M.~J., Ransom, M.~R., and Hilmer, C.~E. (2015).
\newblock Fame and the fortune of academic economists: How the market rewards
  influential research in economics.
\newblock {\em Southern Economic Journal}, 82(2):430--452.

\bibitem[Hirsch, 2005]{hirsch2005index}
Hirsch, J.~E. (2005).
\newblock An index to quantify an individual's scientific research output.
\newblock {\em Proceedings of the National academy of Sciences of the United
  States of America}, 102(46):16569.

\bibitem[Hollis, 2001]{hollis2001co}
Hollis, A. (2001).
\newblock Co-authorship and the output of academic economists.
\newblock {\em Labour economics}, 8(4):503--530.

\bibitem[Kuld and O\'~Hagan, 2018]{kuld2018rise}
Kuld, L. and O\'~Hagan, J. (2018).
\newblock Rise of multi-authored papers in economics: Demise of the ‘lone
  star’ and why?
\newblock {\em Scientometrics}, 114(3):1207--1225.

\bibitem[Laband and Tollison, 2000]{laband2000intellectual}
Laband, D.~N. and Tollison, R.~D. (2000).
\newblock Intellectual collaboration.
\newblock {\em Journal of Political economy}, 108(3):632--662.

\bibitem[Liebowitz, 2014]{liebowitz2014willful}
Liebowitz, S.~J. (2014).
\newblock Willful blindness: The inefficient reward structure in academic
  research.
\newblock {\em Economic Inquiry}, 52(4):1267--1283.

\bibitem[Marchant, 2009a]{marchant2009axiomatic}
Marchant, T. (2009a).
\newblock An axiomatic characterization of the ranking based on the h-index and
  some other bibliometric rankings of authors.
\newblock {\em Scientometrics}, 80(2):325--342.

\bibitem[Marchant, 2009b]{marchant2009score}
Marchant, T. (2009b).
\newblock Score-based bibliometric rankings of authors.
\newblock {\em Journal of the Association for Information Science and
  Technology}, 60(6):1132--1137.

\bibitem[McDowell and Smith, 1992]{mcdowell1992effect}
McDowell, J.~M. and Smith, J.~K. (1992).
\newblock The effect of gender-sorting on propensity to coauthor: Implications
  for academic promotion.
\newblock {\em Economic Inquiry}, 30(1):68--82.

\bibitem[Medoff, 2003]{medoff2003collaboration}
Medoff, M.~H. (2003).
\newblock Collaboration and the quality of economics research.
\newblock {\em Labour Economics}, 10(5):597--608.

\bibitem[Perry and Reny, 2016]{perry2016count}
Perry, M. and Reny, P.~J. (2016).
\newblock How to count citations if you must.
\newblock {\em American Economic Review}, 106(9):2722--41.

\bibitem[Perry and Reny, 2014]{perry2014count}
Perry, M. and Reny, P.~J. (June, 2014).
\newblock How to count citations if you must*.
\newblock {\em Working Paper}, pages 1--25.

\bibitem[Price, 1981]{price1981multiple}
Price, D.~S. (1981).
\newblock Multiple authorship.
\newblock {\em Science}, 212(4498):986--986.

\bibitem[Quesada, 2011]{quesada2011axiomatics}
Quesada, A. (2011).
\newblock Axiomatics for the hirsch index and the egghe index.
\newblock {\em Journal of Informetrics}, 5(3):476--480.

\bibitem[Sartwell, 2017]{sart}
Sartwell, C. (2017).
\newblock Beauty, the stanford encyclopedia of philosophy.

\bibitem[Sauer, 1988]{sauer1988estimates}
Sauer, R.~D. (1988).
\newblock Estimates of the returns to quality and coauthorship in economic
  academia.
\newblock {\em Journal of Political Economy}, 96(4):855--866.

\bibitem[Shapley, 1953]{shapley1953value}
Shapley, L.~S. (1953).
\newblock A value for n-person games.
\newblock {\em Contributions to the Theory of Games}, 2(28):307--317.

\bibitem[Smart and Waldfogel, 1996]{smart1996citation}
Smart, S. and Waldfogel, J. (1996).
\newblock A citation-based test for discrimination at economics and finance
  journals.
\newblock Technical report, National Bureau of Economic Research.

\bibitem[Stephan, 1996]{stephan1996economics}
Stephan, P.~E. (1996).
\newblock The economics of science.
\newblock {\em Journal of Economic literature}, 34(3):1199--1235.

\bibitem[Stigler, 2003]{stigler2003memoirs}
Stigler, G.~J. (2003).
\newblock Memoirs of an unregulated economist.
\newblock {\em University of Chicago Press}.

\bibitem[Szwagrzak and Treibich, 2019]{szwagrzak2015co}
Szwagrzak, K. and Treibich, R. (2019).
\newblock Teamwork and individual productivity.
\newblock {\em Management Science}.
\newblock forthcoming.

\bibitem[Tarnow, 2002]{tarnow2002coauthorship}
Tarnow, E. (2002).
\newblock Coauthorship in physics.
\newblock {\em Science and engineering ethics}, 8(2):175--190.

\bibitem[Woeginger, 2008a]{woeginger2008axiomatic2}
Woeginger, G.~J. (2008a).
\newblock An axiomatic analysis of egghe’s g-index.
\newblock {\em Journal of Informetrics}, 2(4):364--368.

\bibitem[Woeginger, 2008b]{woeginger2008axiomatic1}
Woeginger, G.~J. (2008b).
\newblock An axiomatic characterization of the hirsch-index.
\newblock {\em Mathematical Social Sciences}, 56(2):224--232.

\end{thebibliography}

\end{document}